# Modeling of an efficient singlet-triplet spin qubit to photon interface assisted by a photonic crystal cavity

Kui Wu*[a], Sebastian Kindel*[b], Thomas Descamps[b], Tobias Hangleiter[b], Jan Christoph Müller[b], Rebecca Rodrigo[a], Florian Merget[a], Hendrik Bluhm[b], and Jeremy Witzens[a]
[a]Institute of Integrated Photonics, RWTH Aachen University, Aachen, 52074, Germany
[b]Quantum Technology Group, 2nd Institute of Physics, RWTH Aachen University, Aachen, 52074, Germany

* These authors have contributed equally to this paper.

## ABSTRACT

Efficient interconnection between distant semiconductor spin qubits with the help of photonic qubits would offer exciting new prospects for future quantum communication applications. In this paper, we optimize the extraction efficiency of a novel interface between a singlet-triplet spin qubit and a photonic qubit. The interface is based on a 220 nm thick GaAs/AlGaAs heterostructure membrane and consists of a gate-defined double quantum dot (GDQD) supporting a singlet-triplet qubit, an optically active quantum dot (OAQD) consisting of a gate-defined exciton trap, a photonic crystal cavity providing in-plane optical confinement and efficient out-coupling to an ideal free space Gaussian beam while accommodating the gate wiring of the GDQD and OAQD, and a bottom gold reflector to recycle photons and increase the optical extraction efficiency. All essential components can be lithographically defined and deterministically fabricated on the GaAs/AlGaAs heterostructure membrane, which greatly increases the scalability of on-chip integration. According to our simulations, the interface provides an overall coupling efficiency of 28.7% into a free space Gaussian beam, assuming an $SiO_2$ interlayer filling the space between the reflector and the membrane. The performance can be further increased by undercutting this $SiO_2$ interlayer below the photonic crystal. In this case, the overall efficiency is calculated to be 48.5%.

**Keywords**: quantum computing, optical interface, spin qubit, photonic crystal cavity, gate-defined quantum dot.

## 1. INTRODUCTION

Interconnecting distant stationary qubits with photonic qubits is a pivotal milestone for the quantum Internet [1], where quantum entanglement is distributed between different quantum technology platforms to enable specialized quantum information applications, such as physically secured quantum communication [2-3] and distributed quantum computation [4]. Among the competing qubit hardware platforms, gate-defined quantum dots (GDQDs) offer promising prospects due to their compatibility with standard top-down fabrication techniques and the potential for integrating multiple locally interconnected qubits. In a GDQD, single electrons are trapped by tunable electrostatic potential minima applied to a two-dimensional electron gas (2DEG) with surface metal gates. Moreover, singlet-triplet spin qubits defined by GDQDs enable all electrical control via the exchange interaction and high speed (tens of nanoseconds) manipulation of qubit states [5-6]. Likewise, the sub-nanosecond scale recombination time enables a high emission rate and offers a potential for high optical efficiency even without Purcell enhancement, which is an advantage over other platforms like nitrogen-vacancy centers in diamond [7-10] and ion-traps [11].

GDQD qubits based on gallium arsenide (GaAs) have demonstrated all the key prerequisites for quantum information applications, including qubit initialization, readout, and coherent control [5-6, 12-15]. In contrast to silicon platforms, the direct bandgap of GaAs offers a straightforward avenue for photonic qubit to spin qubit conversion facilitated by direct optical absorption and emission, making it an attractive material for advancing quantum technologies.

However, coherent interfacing between photonic qubits and spin qubits on GDQD systems remains challenging due to the lack of hole confinement in conventional heterostructures, resulting in a loss of correlated photon information. One possible solution to this issue is integrating an optically active quantum dot (OAQD), for example an InAs self-assembled quantum dot (SAQD) or a fully gate-defined electrostatic exciton trap [16], as an intermediary between the photonic qubit and the GDQD spin qubit. A transfer protocol [17] is then applied to adiabatically and coherently tunnel-couple the photogenerated electron to a singlet-triplet spin qubit in a GDQD. Consequently, the OAQDs need to be placed in close proximity to the GDQD to enable transfer of the electron via tunneling from one dot to the other. For concreteness, we describe a fully gate-defined design using an electrostatic trap, but this approach can be transferred to other devices in which highly efficient optical coupling to planar semiconductor structures need to be combined with electrical connectivity, up to roughly micron-scale device sizes. Such an electrostatic exciton trap has the advantage of allowing a fully lithographically defined fabrication process and thus enhanced reproducibility and spatial control, without compromising the transfer protocol's functionality. Further, the exciton energy is controllable via the quantum-confined Stark effect and features a narrow spectral linewidth, as shown in earlier demonstrations [16].

Efficient coupling between the OAQD and an optical fiber is a critical aspect of a functional optical interface. Due to the large refractive index contrast between GaAs and free space, photons need to be emitted in a narrow escape cone to be able to couple to a free space mode, limiting the outcoupling efficiency. Various optical nanostructures, such as micropillar cavities [18], nanowire waveguides [19], microlenses [20-21], and nanophotonic directional couplers [22], have been investigated to increase the outcoupling efficiency. However, these structures have been designed and measured using SAQDs as single photon sources and need to be adapted or are incompatible with an electrostatic exciton trap. An additional challenge not met by these solutions is the routing of electrical contacts. Thus, there is a demand for an efficient optical interface supporting the integration of GDQDs and gate-defined OAQDs.

Photonic crystal cavities (PCC) are widely used in quantum nanotechnology experiments due to their ability to enhance light-matter interaction through the well-known Purcell effect. InAs SAQDs integrated in GaAs photonic crystal cavities have shown controlled spontaneous emission rates [23-24], high quality single photon emission [25], and strong light-matter coupling [25-27]. Moreover, by carefully adjusting the PCC geometry, cavity modes can be tailored, including wavelength, mode profile, and radiation properties [28-29] to meet application requirements. Given the deterministic fabrication of both GDQDs and photonic crystal structures, integrating them in a single device holds promise. Previous studies have reported enhanced photoluminescence in a PCC [30] and optical absorption in a bulls-eye cavity [31] fabricated on a GaAs/AlGaAs quantum well membrane with metal electrodes. In our previous work, we designed an H4 photonic cavity structure with a gate-defined OAQD, demonstrating excellent coupling efficiency to a free space Gaussian beam [32-33].

In this paper, we present a comprehensive design of a novel singlet-triplet spin qubit to fiber interface. Our approach involves using a gate-defined electrostatic exciton trap as an OAQD, placed at the center of a carefully designed photonic crystal cavity, which allows for electrical contacts to be routed to the GDQD and OAQD. The GDQD, OAQD, and PCC are fully lithographically defined in a 220 nm thick GaAs/AlGaAs heterostructure membrane hosting a 2DEG. The cavity is designed in such a way that highly efficient vertical emission is achieved at the working wavelength of the OAQD (823 nm). Four cavity openings are integrated into the cavity structure to enable electrical connections via the 2DEG. They are engineered to feature a mini-stopband [34] centered around 823 nm to maintain the optical confinement of the cavity, since they would otherwise form photonic crystal waveguides [35] allowing the light to leak out. A gold back-reflector is deposited onto the GaAs/AlGaAs membrane over an intermediate $SiO_2$ layer, to coherently recycle photons emitted towards it and enable unidirectional emission. The entire layer stack (GaAs/AlGaAs membrane + $SiO_2$ + gold reflector) is flipped and transferred to a silicon substrate to which it is attached via epoxy, after which the front side electrodes are fabricated and the PCC is etched. The distance between the reflector and the heterostructure membrane is optimized to ensure that the desired vertical emission is constructively enhanced. According to our calculations, this design ensures that more than 50% of the photons emitted by the OAQD are coupled into a narrow free space beam in the perpendicular direction, which in turn has an optical overlap greater than 50% with an ideal Gaussian beam. This design holds promising potential for enabling efficient and scalable quantum information applications with GDQD and photonic crystal cavity integration.

## 2. GDQD AND OAQD STRUCTURE

The detailed geometry of the GDQD, OAQD, and a quantum dot charge sensor (also known as single-electron transistor) is depicted in Figure 1. The electrodes defining these structures are deposited on a 220 nm GaAs/AlGaAs heterostructure membrane. The electrodes are made of Au/Ti wires with a thickness of 9 nm (2 nm Ti and 7 nm Au). The wires of both GDQD and sensor have a width of 30 nm. The OAQD consists of round guard gates (outer diameter of 400 nm) and a central trap gate (central diameter of 116 nm). Electrical voltages are applied to the gates to create local potential minima in the 2DEG in the 20 nm GaAs quantum well, which confine the singlet-triplet spin qubit in the GDQD and the exciton under the trap gate of the OAQD. The wires defining the GDQD and the sensor are only patterned on the top side of the membrane following a validated geometry, while those defining the OAQD are deposited symmetrically on both sides of the membrane to ensure a strongly localized electric field and independent tuning of the exciton trap via the quantum confined Stark effect [36]. A more detailed description of the gate structures can be found in our previous publications [32-33].

To coherently transfer the information from a photonic qubit to a singlet-triplet spin qubit in a GDQD, entanglement between the photo-excited electron and hole must be eliminated. As reported in detail by Joecker *et al.* [17], one possible transfer protocol is described by:

$$\begin{aligned} |\uparrow\circ\rangle|\omega_1, V\rangle &\xrightarrow{\text{photoexc.}} |\uparrow\circ\rangle|\downarrow\Uparrow\rangle \xrightarrow{\text{adiabatic transfer}} |T_0\rangle|\circ\Uparrow\rangle \\ |\uparrow\circ\rangle|\omega_2, V\rangle &\xrightarrow{\text{photoexc.}} |\uparrow\circ\rangle|\uparrow\Downarrow\rangle \xrightarrow{\text{Rabi + ad. transf.}} |S\rangle|\circ\Uparrow\rangle \end{aligned} \quad (1)$$

where the left (right) bra-ket stands for the occupation of the GDQD (OAQD). An electron (hole) spin is represented by a single (double) arrow. The energy of the incident photon is described by the frequency $\omega_{1/2}$, and the vertical (horizontal) polarization by V (H). The polarization is defined relative to the orientation of the magnetic field that is externally applied along a direction within the plane of the quantum well. Vertical (horizontal) indicates that the polarization, which is always parallel to the quantum well, is perpendicular (parallel) to the magnetic field. The transfer process consists of two steps:

(i) Creation of a bound exciton in the OAQD by the absorption of the incident photon in the Voigt configuration, i.e., in the presence of a strong in-plane magnetic field.

(ii) Adiabatic transfer of the photo-excited electron into the GDQD. Coherent transfer of the electron between the OAQD and the GDQD is achieved by adiabatically increasing the detuning between the electronic levels in the two systems. Additionally, a Rabi pulse is utilized to modulate the detuning to establish the singlet state.

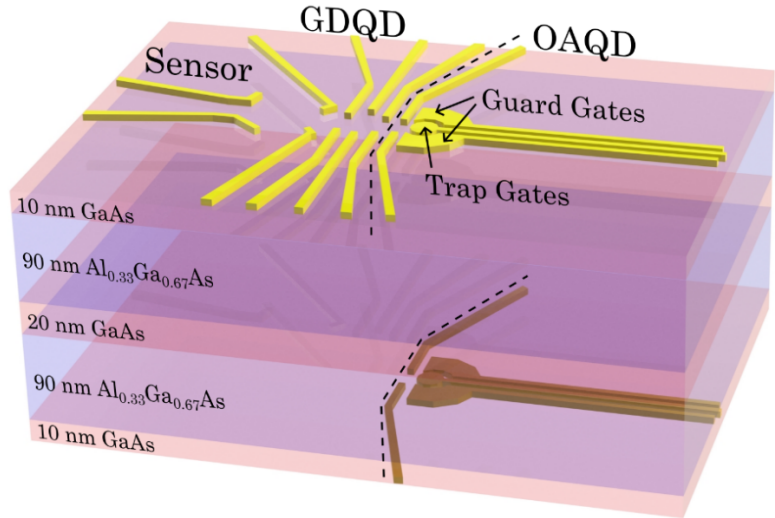

Figure 1. Schematic view of the GDQD, the OAQD, and the quantum dot charge sensor defined on the 220 nm GaAs/AlGaAs heterostructure membrane. The electrodes defining the OAQD are deposited on both sides of the membrane to ensure a strongly localized electric field. On the other hand, the GDQD and the quantum dot charge sensor only require top side electrodes.

By means of the transfer protocol, photonic states with different energies (but the same vertical polarization) are coherently transferred to a single-triplet spin state, wherein the spin-dependent energy of the exciton in the presence of an external magnetic field arises from the different g-factors of electrons and holes. This protocol is estimated to be completed with a fidelity of 84% for a singlet-triplet spin qubit [17], assuming that the capture of the incident photon is successful and a bound exciton is created in the OAQD. Therefore, increasing the conversion efficiency between the photon and the bound exciton is another important aspect, which will be the main topic of this paper. The protocol can be inverted, i.e., starting from a singlet-triplet spin qubit and transferring the quantum information to an energy encoded photon emitted from the OAQD and collected by a single mode fiber. In this paper, we also focus on this reverse scheme as it could be more straightforwardly simulated with the available numerical tools, but conclusions apply to both processes due to reciprocity.

## 3. OPTICAL INTERFACE ASSISTED BY A PHOTONIC CRYSTAL CAVITY

Figure 2(a) presents the schematic overview of our optical interface. To efficiently couple the emitted photon to a single mode fiber, we position the OAQD at the center of a carefully designed 2D PCC. The triangular lattice of holes is etched through the GaAs/AlGaAs heterostructure membrane. Four cavity openings are integrated into the PCC to facilitate electron transport as required by the quantum dot charge sensor and the GDQD without compromising optical confinement at the target wavelength (823 nm). Routing of the Au/Ti wires required for applying the electrostatic potentials and the control pulses through the photonic crystal necessitates a high alignment accuracy of < 15 nm between the PCC and the metallic wires in an electron beam lithography process, considering the width of the wires and the distance between adjacent holes. As demonstrated in our previous publication [33] and in experiments reported by other research groups [37-38], the desired accuracy level can be achieved. A 200 nm thick gold reflector (not shown in Figure 2(a)) is positioned 305 nm below the GaAs quantum well layer to recycle photons emitted downwards. The space between the gold reflector and the GaAs/AlGaAs membrane is filled with an $SiO_2$ interlayer. The reflector position is optimized to maximize the extraction efficiency and ensure a high optical overlap of the far field emission pattern with an ideal Gaussian beam propagating perpendicularly to the membrane in the positive $z$-direction. In addition to enhancing the center lobe of the far field emission of the PCC, the reflector also suppresses side lobes, increasing the purity of the emitted Gaussian beam [32].

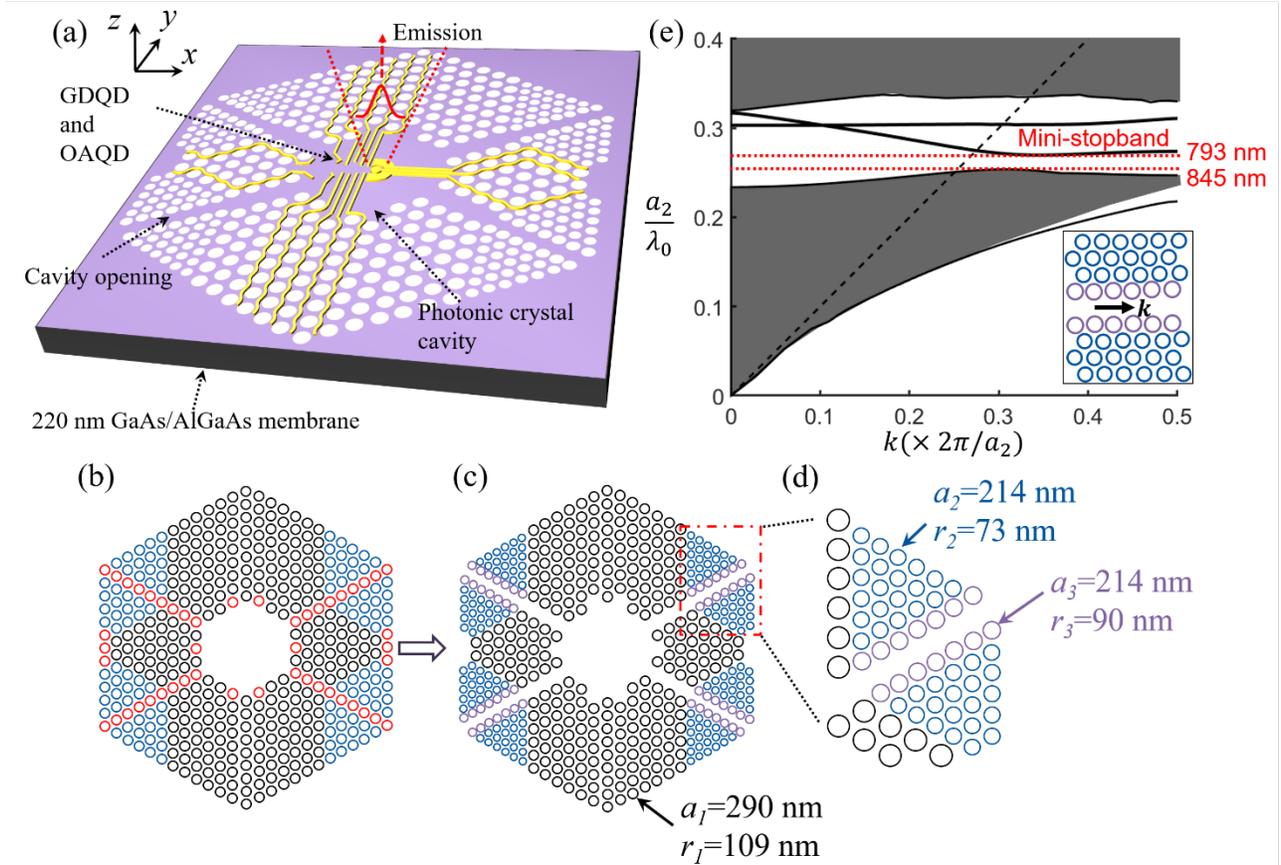

Figure 2. (a) Schematic overview of the optical interface including the photonic crystal cavity, the electrode system, and the cavity openings forming unwanted photonic crystal waveguides in which light propagation is blocked by a mini-stopband. A 200 nm thick gold reflector (not shown) is attached to the membrane via an $SiO_2$ interlayer, whose thickness is optimized to coherently recycle photons emitted downwards. (b) Schematic view of a complete triangular H4 cavity. The holes marked by red circles are removed, while the holes shown in blue are downscaled to move the mini-stopbands to the target wavelength. (c) Final cavity structure after the modification. (d) Detailed view of the lattice around the openings. The holes immediately adjacent to the openings have an enlarged radius in order to achieve a wider mini-stopband. (e) Calculated band structure of the cavity opening from (d). The dashed black line indicates the light line in vacuum. A mini-stopband opens from $\lambda_0 = 793$ nm to $\lambda_0 = 845$ nm. The inset shows the calculated structure and the direction of the in-plane wave-vector.

We design the PCC by modifying a hexagonal H4 cavity [32], with a lattice constant $a_1$ = 290 nm and a hole radius $r_1$ = 109 nm (Figure 2(a)). Those parameters ensure that a transverse electrical (TE) photonic bandgap opens from 717 nm to 1021 nm for a vacuum suspended planar lattice etched in the GaAs/AlGaAs membrane. We remove the holes marked in red in Figure 2(b) for two reasons. First, this simplifies the routing of metal wires and allows for optimal positioning to minimize optical absorption. Second, the openings in the diagonal directions enable a continuous 2DEG within that region of the quantum well, which is essential for the functionality of the quantum dot charge sensor and the GDQD. The reason underlying this need is that the surface charge on the side walls of the etched holes depletes the free carriers in their vicinity, so that electron transport is suppressed in the regular photonic crystal lattice. We have measured the electrical resistance between two ohmic contacts, blocked by a comparable number of layers of holes with the same lattice pitch and hole radius as the photonic crystal in Figure 2(a), to be around 1 MΩ. However, this resistance drops to around 4 kΩ, if an opening similar to the cavity openings in Figure 2(a) is created. This is sufficiently small not to additionally burden electron transport, given that the measured resistance already reaches several kΩ before etching the PCC.

On the other hand, the cavity openings support undesired photonic crystal waveguide modes, which destroy the optical confinement of the PCC. We remedy this issue by utilizing the mini-stopband of the so-formed photonic crystal waveguides [34], which is a result of the anti-crossing of different waveguide modes with the same symmetry. According to our simulations, such a mini-stopband opens from 1030 nm to 1058 nm after the removal of the red holes in Figure 2(b). In order to shift the mini-stopband to the working wavelength of the OAQD, the holes marked in blue in Figure 2(b) are downscaled to a lattice pitch $a_2 = a_3$ = 214 nm, as indicated by the blue and purple circles in Figure 2(c) and 2(d). The downscaled holes have a radius of $r_2$ = 73 nm (marked in blue), while the holes closest to the openings (marked in purple) have an increased radius of $r_3$ = 90 nm. This last modification allows for a wider mini-stopband without burdening the fabrication process further. Figure 2(e) shows the TE band structure calculated along the direction of the cavity opening. As we can see, the mini-stopband is shifted to the wavelength range from 793 nm to 845 nm with an increased spectral width of 52 nm. The shaded areas represent the extended modes in the downscaled crystal (lattice constant $a_2$ and hole radius $r_2$). The guided waveguide modes are indicated by the black solid lines. The OAQD's working wavelength (823 nm) falls within the center of this mini-stopband, achieved through parameter optimization for rapid decay of the electromagnetic field along the openings.

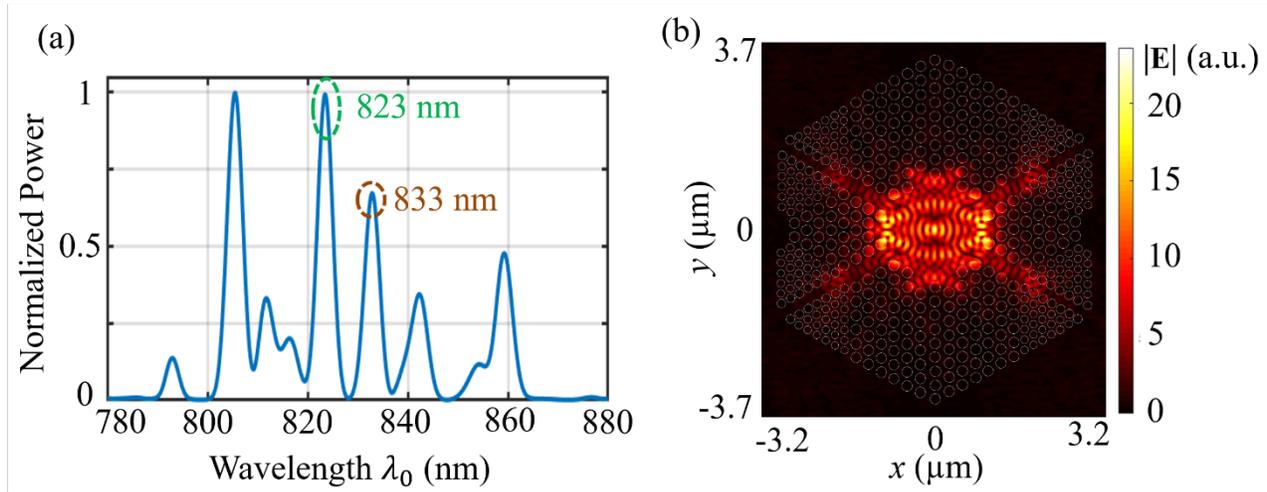

Figure 3. (a) Normalized cavity power spectrum from 780 nm to 880 nm obtained from a cavity ring-down simulation. The target mode, which has a high extraction efficiency, is marked by a green dashed circle at 823 nm. For comparison to the target mode, another cavity mode at 833 nm is also highlighted by a brown circle. (b) Real space cavity mode profile of the 823 nm target mode excited by a y-dipole. An exponentially decaying electric field is observed along the cavity openings as a consequence of the mini-stopband.

We performed three-dimensional finite-difference time-domain (3D FDTD) simulations to evaluate the theoretical performance of the complete structure by using a commercially available software package (Lumerical FDTD). Since the polarization of the photon emitted by the OAQD by pure excitonic states is determined by the orientation of the in-plane magnetic field in the Voigt configuration, as a consequence of the exciton state energy splitting, we model the OAQD by an electric dipole orientated in the *y*-direction (V) with the in-plane magnetic field oriented along the *x*-direction. The bandgap energies of GaAs and Al$_{0.33}$Ga$_{0.67}$As under liquid helium cryogenic temperature (4 K) are $E_{\text{4K, GaAs}}$ = 1.52 eV and $E_{\text{4K, AlGaAs}}$ = 1.90 eV [39], corresponding to optical transition wavelengths of 816 nm and 652 nm, respectively. Consequently, Al$_{0.33}$Ga$_{0.67}$As is fully transparent at 823 nm, while GaAs exhibits weak absorption at this wavelength due to the Urbach tail [40] and exciton absorption near the band edge. However, as our 220 nm membrane structure has a small GaAs fraction (two 10 nm cap layers and a 20 nm quantum well layer), we modeled the GaAs layers as lossless for simplicity. The refractive indices of GaAs and Al$_{0.33}$Ga$_{0.67}$As are estimated to be $n_{\text{4K, GaAs}}$ = 3.59 and $n_{\text{4K, AlGaAs}}$ = 3.38 at 823 nm at 4 K, considering their temperature dependence [41] and linearly extrapolating them. The inaccuracy in the assumed refractive indices introduced by the linear extrapolation will be addressed by experimental iterations. For the SiO$_2$ interlayer, we used an index of $n_{\text{SiO2}}$ =1.45.

The cavity power spectrum obtained from a cavity ring-down simulation is presented in Figure 3(a) from 780 nm to 880 nm. The cavity is excited by an electric dipole oriented along the *y*-direction and located at the center of the cavity, which corresponds to the position of the OAQD. Due to the large size of the cavity, chosen to minimize electrostatic interaction between the quantum dots and surface charges at the etched holes, multiple resonant peaks are observed. The target mode at 823 nm is specifically highlighted with a dashed green circle. The parameters of the PCC, including the lattice constants and hole radii, are carefully engineered to ensure that the wavelength of this target mode coincides with the OAQD's working wavelength at 823 nm.

In Figure 3(b), the cavity field profile at 823 nm is plotted in the plane of the PCC. It is evident that the optical confinement is well maintained, thanks to the presence of the mini-stopband. The electromagnetic field decays exponentially as it propagates along the openings. This feature is essential for guiding the emitted photons upwards.

Pronounced vertical radiation is achieved for the target mode at 823 nm. Figure 4(a) shows the projected electrical field intensity at a distance of 1 m away from the membrane, obtained by decomposing the near field into a series of plane waves and applying a far-field transform. The targeted single-beam vertical emission is observed, clearly indicated by the pronounced far field intensity peak at $x$ = 0 and $y$ = 0. The 200 nm thick gold reflector is positioned 305 nm below the OAQD, which corresponds to an SiO$_2$ interlayer thickness of 195 nm. The reflector position is optimized to facilitate constructive interference and enhance the central emission peak, while simultaneously suppressing the side lobes [32].

The pronounced vertical emission observed in the target mode is a result of Bragg-scattering caused by the periodic lattice of the PCC. Figure 4(b) shows the cavity mode profile in Fourier space for the target mode. For a triangular lattice, the reciprocal lattice (white dots) is spanned by two primitive lattice vectors

$$\vec{V}_1 = \frac{2\pi}{a_1}\vec{e}_y + \frac{2\pi}{\sqrt{3}a_1}\vec{e}_x, \qquad \vec{V}_2 = \frac{2\pi}{a_1}\vec{e}_y - \frac{2\pi}{\sqrt{3}a_1}\vec{e}_x \qquad (2)$$

where $a_1$ = 290 nm is the lattice constant of the PCC, and $\vec{e}_{x/y}$ are unit vectors in the reciprocal space. The target mode exhibits dominant *k*-space components around the reciprocal lattice points at $\pm(\vec{V}_1 - \vec{V}_2)$, which form a standing wave along the *x*-direction. Enhanced vertical emission is therefore achieved through Bragg-scattering of these dominant *k*-space components to the $\Gamma$-point ($k_x = k_y = 0$). Additionally, standing waves along the directions of the cavity openings are clearly visible in Figure 4(b).

For comparison to the pronounced vertical emission observed from the target mode at 823 nm, the far field emission pattern and the *k*-space profile of the 833 nm cavity mode are depicted in Figure 4(c) and 4(d). We can clearly see that the dominant field components of this mode are not at the reciprocal lattice points of the PCC and thus cannot get efficiently coupled to the Γ-point. This results in the complex far-field emission pattern observed in Figure 4(c), that would couple poorly to a single mode fiber.

To calculate the probability of photons emitted by the OAQD to couple into a single mode fiber, we apply the following method. First, the electromagnetic field is recorded by a 2D monitor parallel to the membrane, with dimensions of 5 μm by 5 μm and located at a vertical distance of $z$ = 130 nm above the OAQD in free space (20 nm above the upper surface of the membrane). The electromagnetic power $P_r$ is calculated by integrating the normal component of the Poynting vector over the entire monitor surface. We define the radiation efficiency

$$\eta_r = \frac{P_r}{P_{\text{dipole}}} \qquad (3)$$

as the fraction of dipole power $P_{\text{dipole}}$ that is transmitted through the monitor surface $P_r$. This radiation efficiency $\eta_r$ corresponds to the probability that the photons emitted by the OAQD escape the membrane via the top surface and propagate into free space.

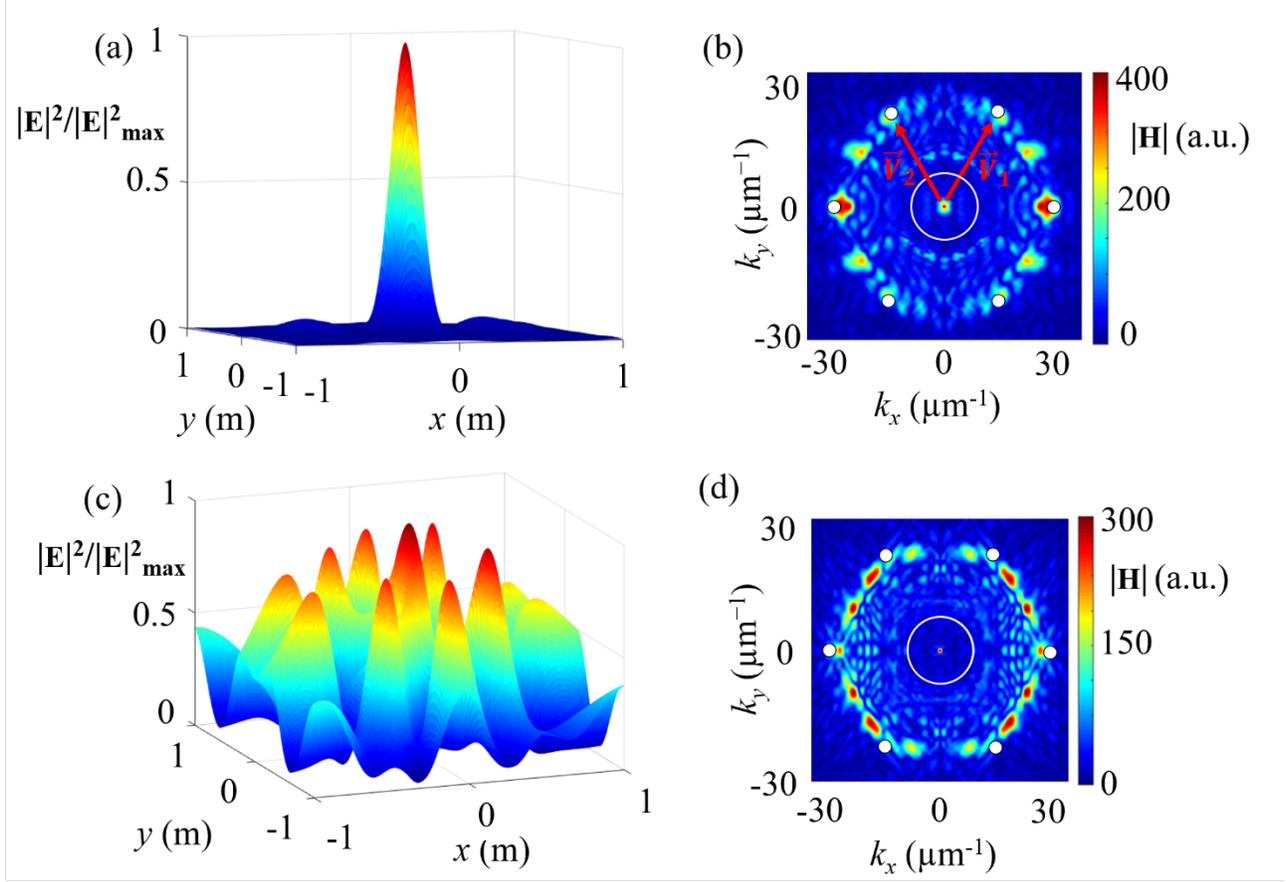

Figure 4. (a) Far field pattern at 823 nm calculated on a plane parallel to the membrane surface. The plane has a size of 2 m by 2 m in the $x$ and $y$-directions and is located at a distance $z = 1$ m away from the OAQD ($x, y, z = 0$). Vertical emission with a pronounced Gaussian shape is observed. (b) Fourier space distribution of the target PCC mode. The red arrows indicate the primitive reciprocal lattice vectors and the white circle represents the light line in free space. The dominant $k$-space components are located around the reciprocal lattice points at $\pm(\vec{V}_1 - \vec{V}_2)$, which results in the enhanced vertical emission. (c) Far field pattern at 833 nm, for the other mode marked in Figure 3(a) for comparison. The complex pattern is very different from a Gaussian beam profile. Note that the far field patterns are independently normalized in (a) and (c). (d) Fourier space distribution of the PCC mode at 833 nm. The dominant field components do not coincide with the reciprocal lattice points.

Next, we calculate the optical overlap (OV) between the electromagnetic field recorded by the monitor and an ideal Gaussian beam using the formula [42]:

$$\text{OV}(\theta) = \frac{|\iint (\vec{E}_m \times \vec{H}_g^*(\theta) + \vec{E}_g^*(\theta) \times \vec{H}_m) \cdot d\vec{S}|^2}{4 \iint Re(\vec{E}_m \times \vec{H}_m^*) \cdot d\vec{S} \iint Re(\vec{E}_g(\theta) \times \vec{H}_g^*(\theta)) \cdot d\vec{S}} \quad (4)$$

Here, $\vec{E}_m$ ($\vec{H}_m$) is the electric (magnetic) field recorded by the monitor, and $\vec{E}_g(\theta)$ ($\vec{H}_g(\theta)$) represents the electric (magnetic) field of a linearly polarized Gaussian beam with an $1/e^2$ intensity half angle $\theta$. To generate the Gaussian beam numerically, we construct a series of plane waves in momentum space, taking into account the Gaussian amplitude distribution and their polarization relative to the dipole moment oriented in the $y$-direction. We then apply an inverse Fourier transform to calculate the real space beam profile. The spatial position of the Gaussian beam is optimized separately to maximize OV.

Finally, we define the overall efficiency of our optical interface as:

$$\eta(\theta) = \eta_r \text{OV}(\theta) \quad (5)$$

which is equal to the probability that the emitted photons couple into a free space Gaussian beam with a divergence angle $\theta$, which can be straightforwardly coupled to the fundamental mode of a single mode fiber with the help of a lens system.

OV and $\eta(\theta)$ are calculated for the two cavity modes highlighted in Figure 3(a) at 823 nm and 833 nm and plotted in Figure 5. For the 823 nm target mode, a maximum optical overlap ($\text{OV}_{max}$) of 0.526 is achieved at a small divergence angle of 10 degrees due to the pronounced vertical emission with a radiation efficiency of $\eta_r = 54.6\%$. As a result, the maximum overall efficiency ($\eta_{max}$) is calculated as $\eta_{max} = \eta_r \text{OV}_{max} = 28.7\%$. On the other hand, the optical overlap curve for the 833 nm mode shows a continuous increase with the divergence angle up to the maximum investigated angle of 80 degrees and stays below $\text{OV}_{max} = 0.101$. This results in $\eta_{max} = 3.8\%$ with $\eta_r = 37.2\%$. This mode lacks the pronounced vertical emission characteristic observed from the 823 nm mode, resulting in poor overall efficiencies even at higher divergence angles and serves to highlight the importance of careful cavity mode design.

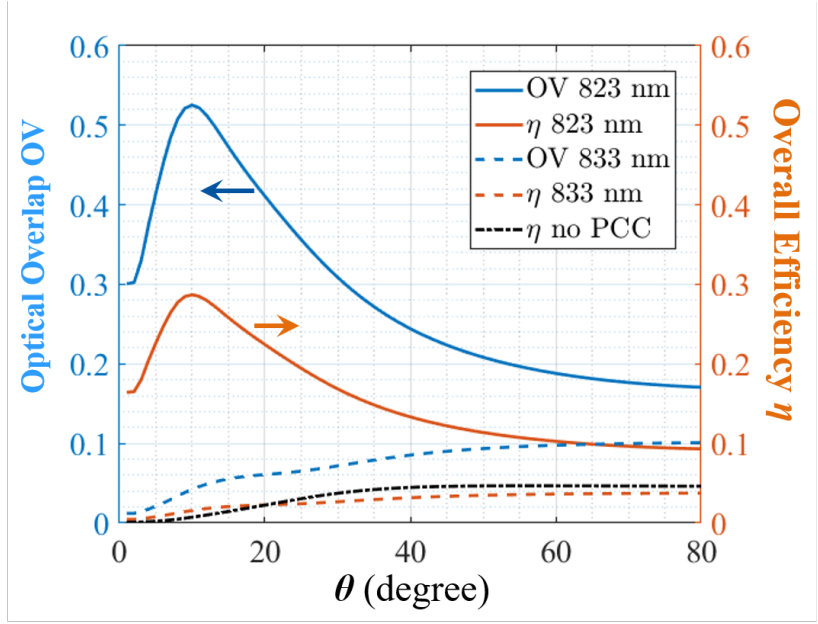

Figure 5. Calculated OVs (blue curves) and overall efficiencies (orange curves) as a function of the divergence angle of a Gaussian beam for the two highlighted cavity modes at 823 nm (solid curves) and 833 nm (dashed curves). As a comparison, the overall efficiency without the PCC is also plotted (black dotted curve).

To demonstrate the effectiveness of the PCC, we also calculate the overall efficiency when the PCC is removed, i.e., an unetched GaAs/AlGaAs membrane with the same gate structure, SiO$_2$ interlayer, and gold reflector at the same distance. In this case, we obtain $\text{OV}_{max} = 0.600$ for a divergence angle of 54 degrees, even slightly better than the PCC, but a much reduced overall $\eta_{max} = 4.6\%$ resulting from the poor radiation efficiency of $\eta_r = 7.8\%$ in absence of the PCC. The function of the PCC can thus be summarized as enhancing the radiation efficiency, while at the same time maintaining a Gaussian emission profile. Moreover, since the PCC significantly reduces the divergence angle of the emitted Gaussian beam, an objective with a lower numerical aperture and thus with a longer working distance can be used down the line to pick up the beam, facilitating the experimental implementation.

Figure 6(a) presents a comparison of the calculated OV and $\eta$ for the PCC structure with and without the SiO$_2$ interlayer. The suspended PCC structure (without the interlayer), which could be fabricated by undercutting the SiO$_2$ in a final fabrication step, is represented by the dashed curves, while the PCC with the SiO$_2$ interlayer is depicted by the solid curves. In both cases, the distance between the gold reflector and the GaAs quantum well layer is optimized to achieve constructive interferences for the central Gaussian peak, with distances of 305 nm and 410 nm from the quantum well layer for the PCC with and without the interlayer, respectively. We see in Figure 6(a) that the maximum overlap $\text{OV}_{max}$ is 0.525 (0.839) with (without) the interlayer at the same divergence angle $\theta = 10°$. The calculated radiation efficiency is $\eta_r = 54.6\%$ with the interlayer and $\eta_r$ = 57.8% without the interlayer. Therefore, the maximum overall efficiencies with and without the interlayer are $\eta_{max} = \eta_r OV_{max}$ = 28.7% and 48.5%, respectively. This comparison shows that the SiO$_2$ interlayer in the overall PCC structure leads to reduced OV and $\eta_r$. This is due on the one hand to increased emission towards the reflector resulting from the decreased refractive index contrast. Additionally, the presence of the SiO$_2$ interlayer breaks the symmetry of the structure along the z-direction and induces coupling between the quasi-transverse-electric (TE) and quasi-transverse-magnetic (TM) modes [43]. Since the bandgap is open for the TE fields only, this contributes further to the reduction of the overall emission efficiency in the following way:

The photonic crystal cavity ($a_1$ = 290 nm, $r_1$ = 109 nm) supports a TE bandgap from 717 nm to 1021 nm. Additionally, the cavity opening has a TE mini-stopband from 793 nm to 845 nm, as illustrated in Figure 2(e). Therefore, no TE mode propagating out of the cavity exists in the wavelength range from 793 nm to 845 nm and TE cavity modes are observed within the photonic crystal cavity. However, for the TM modes, such a bandgap does not exist. When the reflector symmetry is broken by the asymmetric SiO$_2$ interlayer in the z-direction and TE-TM coupling is possible, the TE-like cavity modes can couple into the TM-like propagating photonic crystal modes or the TM-like propagating waveguide modes and escape the cavity. This results in an additional in-plane loss inside the semiconductor membrane. Figure 6(b) shows the k-space profile of the target cavity mode together with the equifrequency contours of the TM-like photonic crystal modes at the same wavelength (823 nm), represented by red dotted curves. The presence of field components on the equifrequency contours of the TM-like photonic crystal modes suggests the presence of in-plane TE-TM coupling loss.

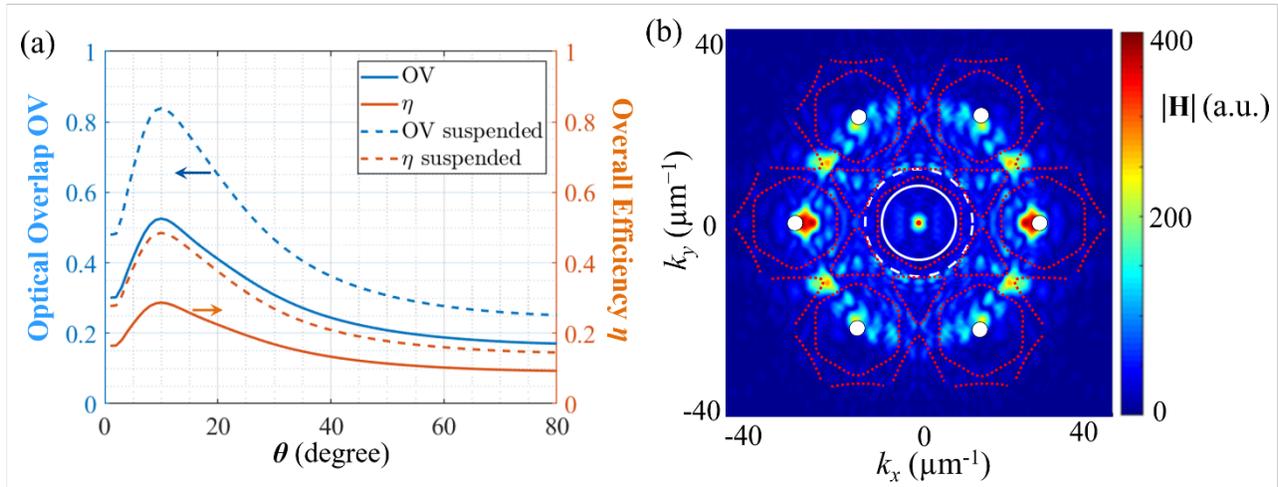

Figure 6. (a) Calculated OV and overall efficiency as a function of $\theta$ for the cavity mode at 823 nm. The reflector position is implicitly optimized for each case. In the presence of the SiO$_2$ interlayer, OV$_{max}$ decreases from 0.839 to 0.526 and the $\eta_{max}$ from 48.5% to 28.7% at $\theta$ = 10° as a consequence of the broken symmetry along the z-direction. (b) Fourier space distribution of the target TE-like cavity mode at 823 nm overlaid with the equifrequency contours of the leaky TM-like photonic crystal modes at the same wavelength. The light line of vacuum (SiO$_2$) is depicted by the white solid (dashed) circle, and the k-vectors of the TM-like photonic crystal modes are indicated by the red dotted contour. An overlap between the TE-like cavity mode field components and the TM-like photonic crystal modes is observed.

In addition to the TE-TM coupling loss, two other loss mechanisms (radiation loss via the bottom interface and optical absorption of the metal gates) also limit the performance of our optical interface. The gold reflector utilized in this setup reflects light traveling in the negative z-direction and efficiently enhances the Gaussian emission towards the top, which is emitted perpendicular to the chip surface. However, photons that are emitted at non-perpendicular angles to the reflector's surface might experience multiple reflections that guide them along the 2D slab formed by the gold reflector and the GaAs/AlGaAs membrane in the xy-plane. As a result, there is a possibility of radiation loss via the bottom semiconductor interface, which cannot be neglected. Light inside the SiO$_2$ light cone but outside the vacuum light cone stays guided. Moreover, even for the light within the vacuum light cone that eventually escapes the slab, this leads to a distortion of the emitted field profile, reducing the overlap OV. Moreover, losses are introduced from multiple reflections from the metal mirror, that induces some absorption.

The optical absorption of the Au/Ti gates introduces another loss mechanism due to the presence of surface plasmon polaritons (SPP) at the metal/vacuum and metal/SiO$_2$ interfaces excited by the evanescent electromagnetic field of the target cavity mode. To reduce the SPP absorption, the gates/wires are carefully routed to ensure that the metal/vacuum and metal/SiO$_2$ interfaces are mostly parallel to the dominant electric field component inside the cavity, which is $E_y$. This orientation is chosen because the dominant electric field component of SPP modes is perpendicular to the metal/dielectric interface [44]. By routing the gates/wires in this manner, the interaction between the SPP and the cavity mode is reduced, leading to decreased optical absorption.

To study the relative importance of the different loss mechanisms, we calculated the quality factors ($Q$) of the PCC structure by exciting the cavity with an $x/y$-dipole at the position of the OAQD and recording the cavity spectrum. The results are summarized in Table 1. The overall $Q_{total}$ is evaluated by fitting the resonance peak via a Lorentzian curve and calculating the full-width-half-maximum (FWHM). We also evaluated the ratio of the lost power to the total power emitted by the dipole $\Gamma$ for the different loss channels and converted this into the corresponding partial quality factors due to gates absorption ($\Gamma_{gates}$ and $Q_{gates}$), TE-TM coupling loss ($\Gamma_{TE-TM}$ and $Q_{TE-TM}$), downward radiation ($\Gamma_{bottom}$ and $Q_{bottom}$), and upward radiation ($\eta_r$ and $Q_{top}$). Note that we use the already introduced radiation efficiency $\eta_r$ instead of the notation $\Gamma_{top}$ to represent the ratio of the power emitted through the top interface. A high emission efficiency corresponding to a high $\eta_r$ is obtained if $Q_{top}$ is substantially lower than all the other partial Q-factors, which correspond to unwanted loss channels.

|  | $Q_{total}$ | $\Gamma_{gates}$ / $Q_{gates}$ | $\Gamma_{TE-TM}$ / $Q_{TE-TM}$ | $\Gamma_{bottom}$ / $Q_{bottom}$ | $\eta_r$ / $Q_{top}$ | OV$_{max}$ | $\eta_{max}$ |
|---|---|---|---|---|---|---|---|
| SiO$_2$ $y$-dipole | 232 | 25.8% / 899 | 3.5% / 6620 | 15.9% / 1456 | 54.7% / 424 | 0.526 | 28.7% |
| SiO$_2$ $x$-dipole | 161 | 30.9% / 521 | 6.6% / 2426 | 22.5% / 716 | 39.4% / 409 | 0.347 | 13.7% |
| Suspended $y$-dipole | 293 | 34.4% / 851 | 1.4% / 20285 | 5.9% / 4996 | 57.8% / 507 | 0.839 | 48.5% |
| Suspended $x$-dipole | 236 | 42.5% / 555 | 2.3% / 10301 | 7.8% / 3026 | 46.3% / 510 | 0.719 | 33.3% |

Table 1. Calculated power ratio $\Gamma$ and partial quality factors for each loss channel. Four different schemes are investigated: PCC with SiO$_2$ interlayer and excited by a $y$-dipole, PCC with SiO$_2$ interlayer and excited by an $x$-dipole, vacuum suspended PCC excited by a $y$-dipole, and vacuum suspended PCC excited by an $x$-dipole. The distance between the reflector and the GaAs/AlGaAs membrane is optimized independently for all 4 cases. The polarization of the Gaussian beam used for calculating the OV is always in the direction of the dipole moment.

We see from Table 1 that the partial quality factor of the metal gates ($Q_{gates}$) is highly dependent on the dipole orientation. For an $x$-dipole, $Q_{gates}$ is calculated as 521 and 555 with and without the SiO$_2$ interlayer, respectively. In contrast, the same $Q_{gates}$ increases to 899 and 851 for a $y$-dipole due to reduced SPP absorption as a result of the optimized gate geometry. Moreover, $Q_{TE-TM}$ is significantly correlated to the presence or absence of the interlayer. When an SiO$_2$ interlayer is used, $Q_{TE-TM}$ is reduced by one order of magnitude due to the increased TE-TM coupling loss. Even for a suspended PCC structure, the TE-TM coupling loss is still non-negligible (with a power ratio of 1.4% for the $y$-dipole and 2.3% for the $x$-dipole excitation), which is possible since the gates also break the symmetry in the $z$-direction, as the GDQD is only patterned on the top semiconductor interface. The power loss via the bottom interface is also increased when an SiO$_2$ interlayer is present, because the reduced refractive index contrast enhances the downward emission. We notice that $Q_{top}$ is around 500 for a suspended structure and around 400 for the case of an interlayer. The dipole orientation has minimal influence on $Q_{top}$.

Table 1 also includes the calculation results for the maximum optical overlap (OV$_{max}$) in all four schemes. We notice a pronounced correlation between OV$_{max}$ and $Q_{bottom}$, which may be explained by multiple reflections and partial guiding in the underlying interlayer being applied to the downward emission before it escapes the structure through the top, distorting the emission profile. Based on these observations, undercutting of the membrane appears desirable from a performance perspective, even though it increases the complexity of the fabrication process.

## 4. CONCLUSION

In this paper, we present and numerically analyze the extraction efficiency of an optical interface between a singlet-triplet spin qubit and a photonic qubit, including the GDQD, the OAQD, a charge sensor, and the photonic crystal cavity structure. The entire structure can be fully lithographically defined and deterministically fabricated, which greatly increases the scalability of on-chip integration. The photonic crystal cavity is designed such that the dominant wave vectors of the cavity mode at the working wavelength of the OAQD coincide with the reciprocal lattice vectors of the photonic crystal. As a result of strong Bragg scattering, we obtain enhanced vertical emission with a pronounced central lobe. According to our calculations, our design reaches a radiation efficiency of 54.7% and an optical overlap of 0.526 with a narrow Gaussian beam compatible with low numerical aperture collection optics. As a result, we expect an overall efficiency of 28.7%. The performance can be further increased by removing the SiO$_2$ interlayer between the semiconductor membrane and the gold reflector. In this case, the overall efficiency is estimated to be 48.5% with an optical overlap equal to 0.839 and a radiation efficiency of 57.8%.

## 5. ACKNOWLEDGEMENT

This project is funded by the Deutsche Forschungsgemeinschaft (DFG, German Research Foundation) under Germany's Excellence Strategy – Cluster of Excellence Matter and Light for Quantum Computing (ML4Q) EXC 2004/1 – 390534769.